\documentclass{aa}
\usepackage{graphicx}
\usepackage{txfonts}
\usepackage{natbib}
\bibpunct{(}{)}{;}{a}{}{,}
\newcommand{\ea}{et al.}
\newcommand{\cpd}{\ensuremath{\mathrm{c\,d}^{-1}}}
\newcommand{\kmps}{\ensuremath{\mathrm{km\,s}^{-1}}}
\newcommand{\muhz}{\ensuremath{\mu\mathrm{Hz}}}
\newcommand{\hrd}{HR diagram}

\newcommand{\msun}{\ensuremath{M_\odot}}
\newcommand{\rsun}{\ensuremath{R_\odot}}
\newcommand{\lsun}{\ensuremath{L_\odot}}
\newcommand{\xc}{\ensuremath{X_\mathrm{c}}}
\newcommand{\dov}{\ensuremath{d_\mathrm{ov}}}

\newcommand{\mbyh}{\ensuremath{\mathrm{[M/H]}}}
\newcommand{\teff}{\ensuremath{T_\mathrm{eff}}}
\newcommand{\lteff}{\ensuremath{\log T_\mathrm{eff}}}

\newcommand{\bcma}{\ensuremath{\beta}~CMa}
\newcommand{\bcep}{\ensuremath{\beta}~Cephei}

\begin{document}
%
\title{An asteroseismic study of the $\beta$~Cephei star 
\mbox{$\beta$~Canis Majoris}
\thanks{Based on spectroscopic data collected with the
CORALIE \'echelle spectrograph attached to the 1.2m Swiss Euler
telescope at La Silla, Chile.}}

\author{
A.\ Mazumdar \inst{1,2} 
\and  M.\ Briquet
\thanks{Postdoctoral Fellow of the Fund for Scientific Research, Flanders.}
\inst{,1}
\and M.\ Desmet \inst{1} 
\and C.\ Aerts \inst{1,3}
}

\offprints{A. Mazumdar}

\institute{
Instituut voor Sterrenkunde, Katholieke Universiteit Leuven, 
Celestijnenlaan 200 B, B-3001 Leuven, Belgium\\
\email{anwesh.mazumdar@yale.edu}
\and
Astronomy Department, Yale University, P.O. Box 208101, 
New Haven, CT 06520-8101, USA
\and
Department of Astrophysics, University of Nijmegen, PO Box 9010, 
6500 GL Nijmegen, the Netherlands
}

\date{Received ; accepted }

\abstract
{}
{We present the results of a detailed analysis of 452 ground-based
high-resolution high S/N spectroscopic measurements spread over 4.5
years for $\beta$~Canis Majoris with the aim to determine the
pulsational characteristics of this star, and to use them to derive
seismic constraints on the stellar parameters.}
{We determine pulsation frequencies in the Si\,III 4553~\AA\ line with
Fourier methods. We identify the $m$-value of the modes by taking into
account the photometric identifications of the degrees $\ell$. To this
end we use the moment method together with the amplitude and phase
variations across the line profile. The frequencies of the identified
modes are used for a seismic interpretation of the structure of the
star.}
{We confirm the presence of the three pulsation frequencies already
detected in previous photometric datasets: $f_1 = 3.9793~\cpd$
($46.057~\muhz$), $f_2 = 3.9995~\cpd$ ($46.291~\muhz$) and $f_3 =
4.1832~\cpd$ ($48.417~\muhz$). For the two modes with the highest
amplitudes we unambiguously identify $(\ell_1,m_1) = (2,2)$ and
$(\ell_2,m_2) = (0,0)$. We cannot conclude anything for the third mode
identification, except that $m_3 > 0$.  We also deduce an equatorial
rotational velocity of $31 \pm 5~\kmps$ for the star. We show that the
mode $f_1$ must be close to an avoided crossing. Constraints on the mass
($13.5 \pm 0.5 \msun$), age ($12.4 \pm 0.7$~Myr) and core overshoot
($0.20 \pm 0.05\,H_P$) of \bcma\ are obtained from seismic modelling
using $f_1$ and $f_2$.}
{}


\authorrunning{Mazumdar et al.}
\titlerunning{Asteroseismology of \mbox{$\beta$ Canis Majoris}}

\maketitle
%

\section{Introduction}
\label{sec:intro}

Many breakthroughs have recently been achieved in the field of
asteroseismology of \bcep\ stars. The observation of a few pulsating
modes led to constraints not only on global stellar parameters but also
on the core overshoot parameter and on the non-rigid rotation of several
\bcep\ stars. In particular modelling has been performed for HD~129929
\citep{aerts03a,dupret04} and $\nu$~Eri
\citep{pamyatnykh04,ausseloos04}. Our aim is to add other \bcep\ stars
to the sample of those with asteroseismic constraints.  

The B\,1\,II-III bright \bcep\ star $\beta$~Canis Majoris (HD\,44743,
HR\,2294, $V_\mathrm{mag} = 1.97$) is particularly interesting to study.
Indeed, earlier photometric and spectroscopic data revealed that this
object exhibits multiperiodicity with rather low frequencies in
comparison with the frequencies of other \bcep\ stars, which would
indicate that \bcma\ is either a reasonably evolved star or oscillates
in modes different from the fundamental. 

The variability of \bcma\ has been known for one century and the star
has been extensively studied. We refer to \cite{albrecht08},
\citet{henroteau18}, \citet{meyer34} and \citet{struve50} for the first
spectroscopic measurements of \bcma. Later \citet{shobbrook73} found
three pulsation frequencies from extensive photometric time series. The
same three frequencies were recently confirmed by \citet{shobbrook06}
who analysed photometric measurements of a multisite campaign dedicated
to the star.

\citet{aerts94} collected spectroscopic data in order to identify the
modes of the known frequencies of \bcma. In this paper, we present a
similar analysis but based on a much larger number of spectra and using
the version of the moment method improved by \citet{briquet03}. We then
construct stellar models which show oscillations in accordance with our
unique identification of the modes of $\beta$~Canis Majoris.

The paper is organised as follows. Section~\ref{sec:spec} describes the
results from our spectroscopic observations, including data reduction,
frequency analysis and mode identification. In Sect.~\ref{sec:model} we
present our seismic interpretation of \bcma. We end the paper with a
discussion about our results in Sect.~\ref{sec:concl}.

\section{Spectroscopic results}
\label{sec:spec}

\subsection{Observations and data reduction} 
\label{sec:spec_obs}

Our spectroscopic data were obtained with the CORALIE \'echelle
spectrograph attached to the 1.2m Leonard Euler telescope in La Silla
(Chile). Because the beat period between the two known dominant
frequencies is about 50 days we collected data on a long time span.
Observations were collected during several runs spread over 4.5 years.
The number of observations and the ranges of their Julian Dates are
given in Table~\ref{tab:log}. In total, we gathered 452 spectra during
1692 days.  

\begin{table}
\begin{center}
\caption{
Observing logbook of our spectroscopic observations of \bcma.
\label{tab:log}
}
\begin{tabular}{cccc}
\hline
\hline
Number of    & \multicolumn{2}{c}{JD} \\
observations & \multicolumn{2}{c}{2450000 +}   \\
             & Start & End \\
\hline
22           & 1591  & 1598 \\
26           & 1654  & 1660 \\
70           & 1882  & 1894 \\
62           & 1940  & 1953 \\
60           & 2227  & 2239 \\
27           & 2569  & 2582 \\
\phantom{0}2 & 2624  & 2624 \\
91           & 2983  & 2994 \\
54           & 3072  & 3085 \\
38           & 3271  & 3283 \\
\hline
\end{tabular}
\end{center}
\end{table}

An on-line reduction of the CORALIE spectra, using the INTER-TACOS
software package, is available. For a description of this reduction
process we refer to \citet{baranne96}. We did a more precise correction
for the pixel-to-pixel sensitivity variations by using all available
flatfields obtained during the night instead of using only one
flatfield, as is done by the on-line reduction procedure.  Finally, all
spectra were normalised to the continuum by a cubic spline function, and
the heliocentric corrections were computed. For our study of the
line-profile variability we used the Si\,III triplet around 4567~\AA.
This triplet is very suitable to study \bcep\ stars since the lines are
strong, dominated by temperature broadening, and not too much affected
by blending \citep[see][]{aertsdecat03}.

\subsection{Frequency analysis}
\label{sec:spec_freq}

We performed a frequency analysis on the first three velocity moments
\mbox{$<v^1>$}, \mbox{$<v^2>$} and \mbox{$<v^3>$} \citep[see][for a
definition of the moments of a line profile]{aerts92} of the Si\,III
4553~\AA\ line by means of the program Period04 \citep{lenz05}. For some
\bcep\ stars \citep[see][]{schrijvers04,telting97b,briquet05} a
two-dimensional frequency analysis on the spectral lines led to
additional frequencies compared to the one dimensional frequency search
in integrated quantities such as moments.  Consequently we also tried to
find other frequencies for \bcma\ by means of this latter method.  

\begin{figure}  
\begin{center}
\hfil\resizebox{85mm}{!}{\includegraphics{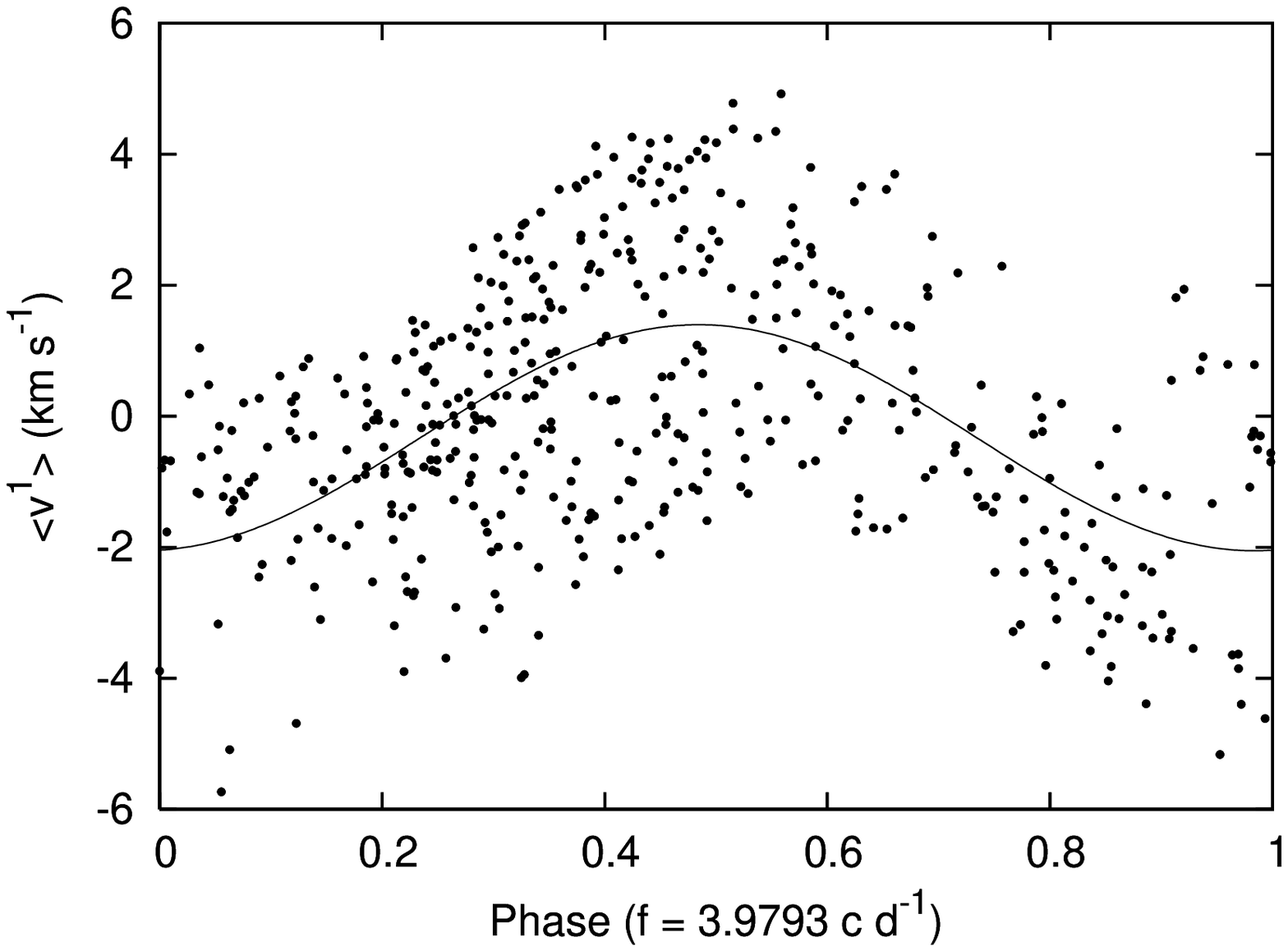}}\hfil
\hfil\resizebox{85mm}{!}{\includegraphics{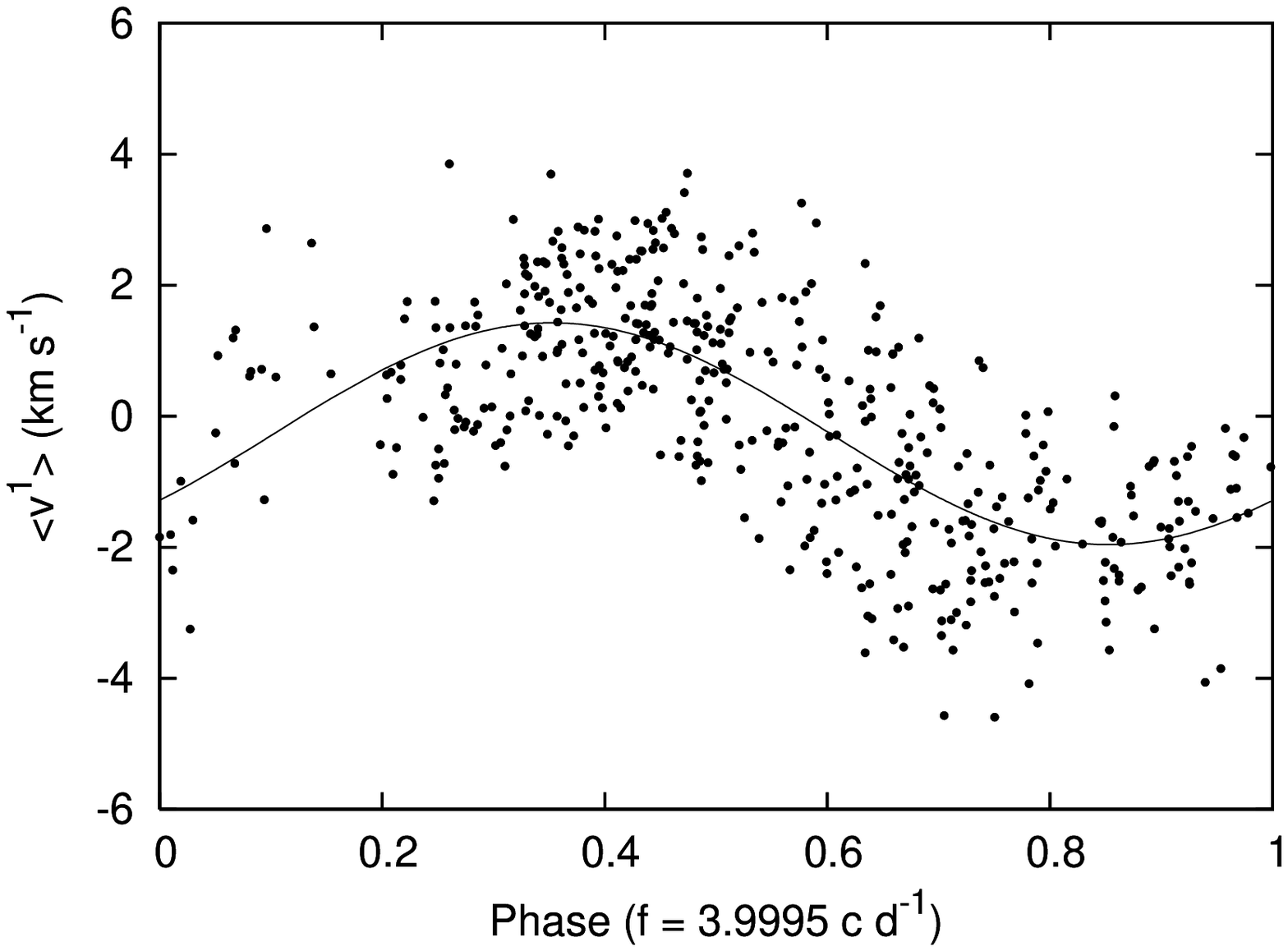}}\hfil
\hfil\resizebox{85mm}{!}{\includegraphics{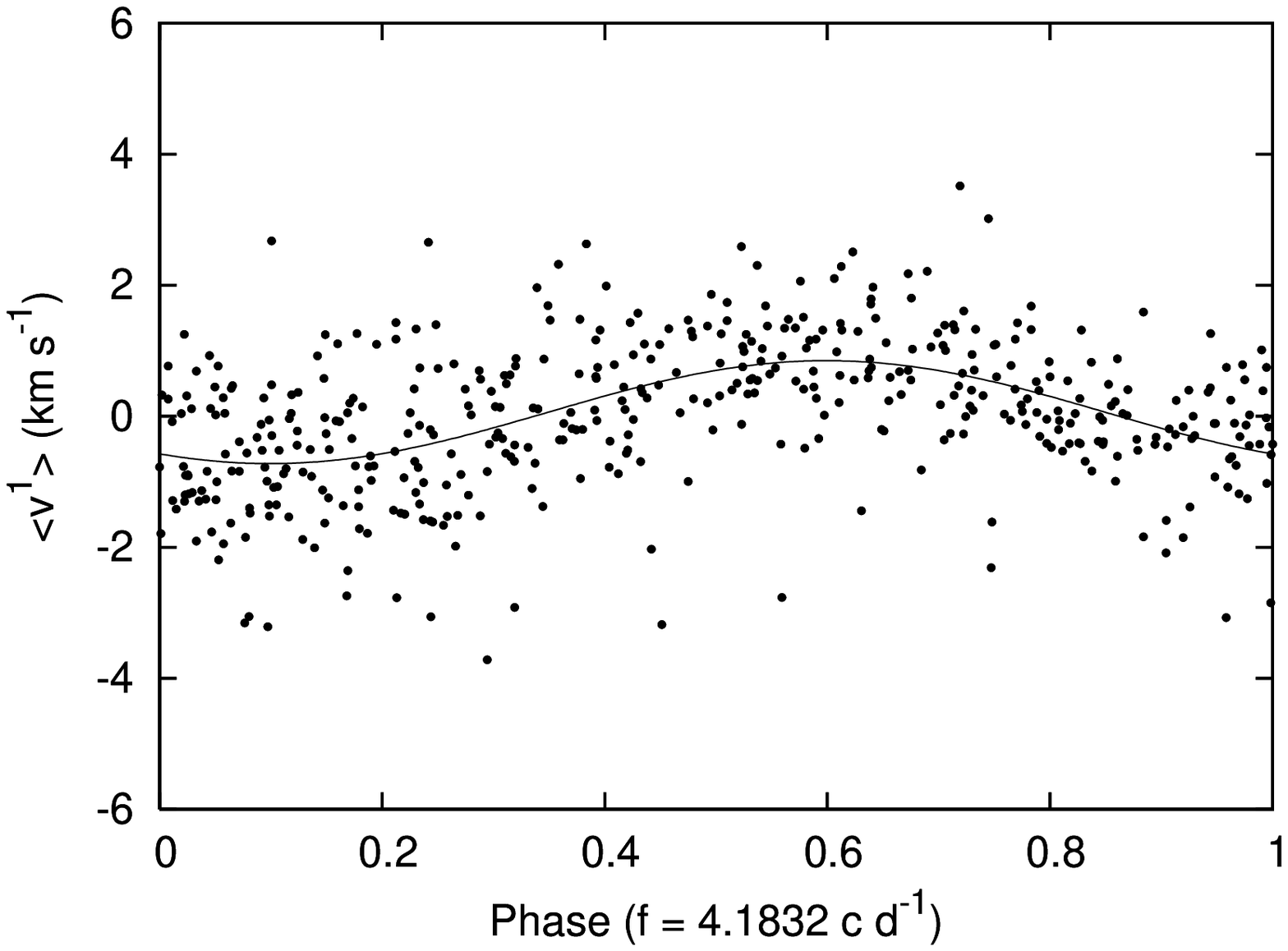}}\hfil
\caption{
From top to bottom: phase diagram of \mbox{$<v^1>$} for $f_1 = 3.9793$
\mbox{c d$^{-1}$}, for $f_2 = 3.9995$ \mbox{c d$^{-1}$} after
prewhitening with $f_1$ and for $f_3 = 4.1832$ \mbox{c d$^{-1}$} after
prewhitening with $f_1$ and $f_2$.  
\label{fig:v1}  
}
\end{center}
\end{figure}  

\begin{table}
\begin{center}
\caption{
Frequencies and amplitudes of the first moment of the Si\,III 4553~\AA\
line together with their standard errors.  The quoted errors for the
frequency are intermediate between the overestimated value given by the
frequency resolution, $1/T \sim 6\times 10^{-4}~\cpd$, and the
underestimated value given by Period04 for an ideal case free from
aliasing and for white uncorrelated noise ($6\times 10^{-6}, 6\times
10^{-6}$ and $2\times 10^{-5}~\cpd$ for $f_1$, $f_2$ and $f_3$
respectively).
\label{tab:fit}
}
\begin{tabular}{ccccc}
\hline
\hline
      & \multicolumn{2}{c}{Frequency}            & Amplitude \\
      & (\mbox{c d$^{-1}$})  & (\muhz)           & (\kmps) \\
\hline
$f_1$ & $3.9793 \pm 0.0001$ & $46.057 \pm 0.001$ & $2.7 \pm 0.1$ \\
$f_2$ & $3.9995 \pm 0.0001$ & $46.291 \pm 0.001$ & $2.6 \pm 0.1$ \\
$f_3$ & $4.1832 \pm 0.0001$ & $48.417 \pm 0.001$ & $0.7 \pm 0.1$ \\
\hline
\end{tabular}
\end{center}
\end{table}

\begin{figure*}
\begin{center}
\includegraphics[width=180mm]{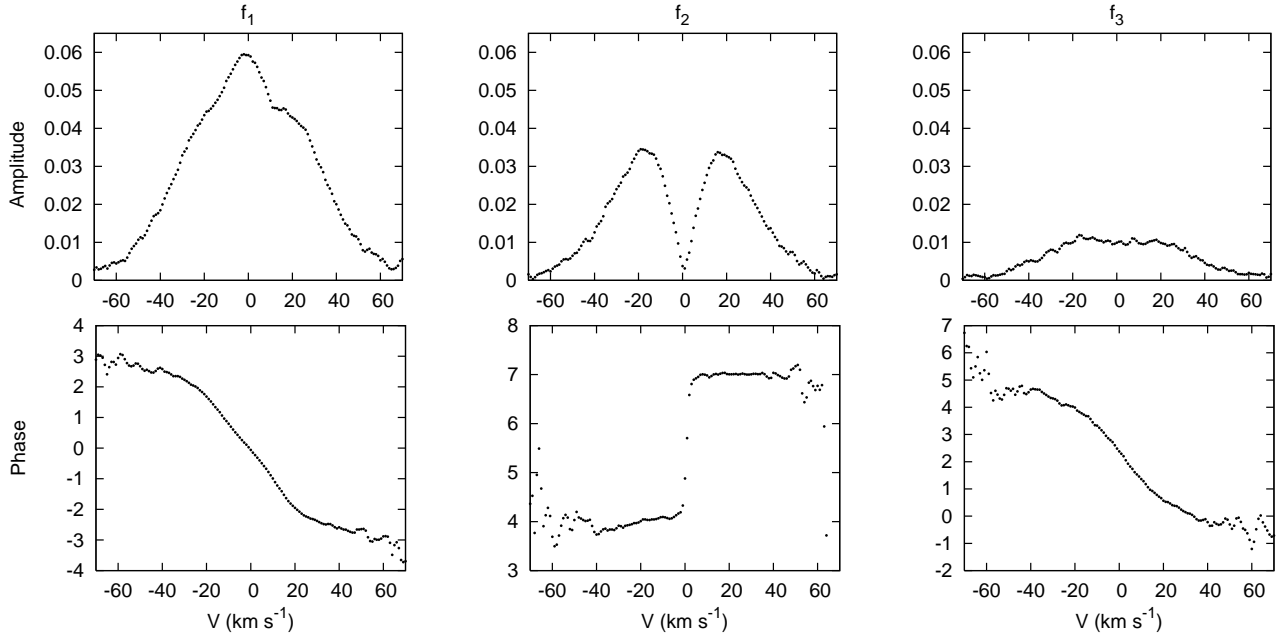}
\caption{
Amplitude and phase distributions for $f_1 = 3.9793~\cpd$, $f_2 =
3.9995~\cpd$ and $f_3 = 4.1832~\cpd$ for the Si\,III 4553~\AA\ line. The
amplitudes are expressed in continuum units and the phases are expressed
in $\pi$~radians.  
\label{fig:amp_phase}  
} 
\end{center}
\end{figure*}

Despite the strong aliasing of our dataset, our frequency analysis
reaffirmed the presence of the already known three pulsating frequencies
of \bcma. $f_1 = 3.9793$ \mbox{c d$^{-1}$}, $f_2 = 3.9995$ \mbox{c
d$^{-1}$} and $f_3 = 4.1832$ \mbox{c d$^{-1}$}
\citep{shobbrook73,shobbrook06}.  Unfortunately no other frequencies
could be discovered in our new spectroscopic data. Phase diagrams of
\mbox{$<v^1>$} for $f_1$, $f_2$ and $f_3$ are shown in
Fig.~\ref{fig:v1}. The frequencies and amplitudes that yielded
to the best fit of \mbox{$<v^1>$} are listed in Table~\ref{tab:fit}.
These three frequencies reduce the standard deviation of the first
moment by 72\%. Fig.~\ref{fig:amp_phase} shows the variations across the
Si\,III 4553~\AA\ line for $f_1$, $f_2$ and $f_3$ \citep[see][for a
definition]{telting97a,schrijvers97}.

We also performed a frequency analysis on the equivalent width but we
could not find any significant frequency. This is not uncommon since
significant EW variations have been found in only a few \bcep\ stars
with photometric variations up to now \citep{deridder02}.

\subsection{Mode identification}
\label{sec:spec_modeid}

Our methodology to identify the modes of \bcma\ is similar to the one
used in \citet{briquet05}, which led to a successful mode identification
for the \bcep\ star \mbox{$\theta$ Ophiuchi}. We refer to that paper for
a detailed explanation on our chosen process.

\subsubsection{Adopted $\ell$-identifications}
\label{sec:spec_modeid_adoptedl}

We make use of spectroscopy to identify the values of the azimuthal
number, $m$, which are not accessible from photometry. In order to limit
the number of parameters in our spectroscopic mode identification we
adopt the degree, $\ell$, as obtained from photometric mode
identification. Recently, \citet{shobbrook06} found $\ell_1 = 2$ for the
first mode, $f_1$, and $\ell_2 = 0$ for the second mode, $f_2$.
However, the degree of the third mode could not be determined.

With our dataset we corroborate the mode with $f_2$ to be radial as
follows. \citet{telting97a} and \citet{schrijvers97} showed that, when
there is a minimum (almost zero) in the amplitude and a corresponding
phase shift of $\pi$ near the centre of the line profile, one can
conclude to be dealing with a radial or a dipole mode. As evident from
Fig.~\ref{fig:amp_phase}, this is clearly the case for $f_2$.

\subsubsection{Moment method}
\label{sec:spec_modeid_moment}

\begin{figure}
\begin{center}
\includegraphics[width=85mm]{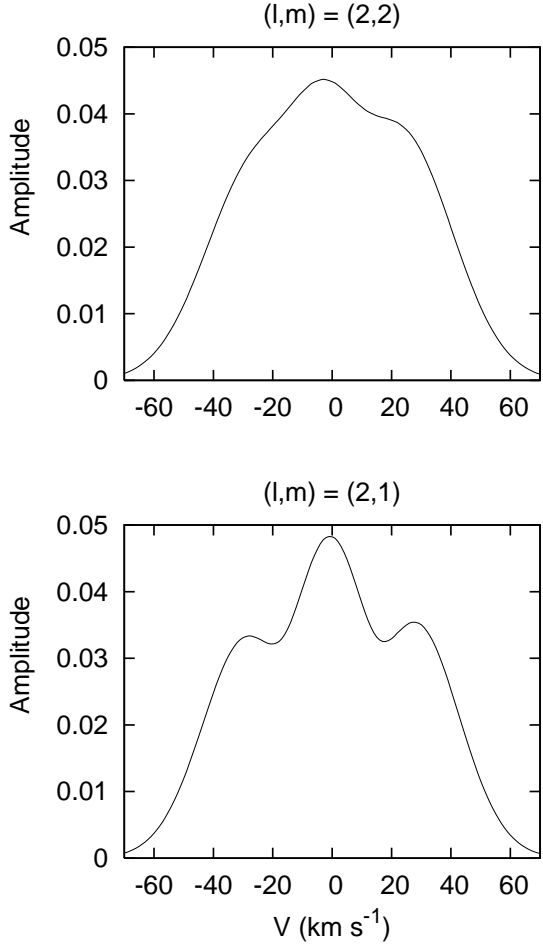}
\caption{
The theoretical amplitude distribution for $f_1$ computed from line 
profile time series generated for the best parameter combinations derived 
with the moment method.
\label{fig:amp_phase_th}  
}
\end{center}
\end{figure}

With the adopted $\ell$-values we then determine the $m$-values by means
of the moment method. The new implementation of this technique was
optimised for multiperiodic signals by \citet{briquet03}. Its
improvements and the huge increase in dataset explain why we obtain a
mode identification different from \citet{aerts94} who used an older
version of this method and whose data did not cover the beat periods of
\bcma. 

The theoretical moment values to be compared to observed moment values
are computed by fixing the following parameters. A linear limb-darkening
coefficient $u$ of 0.292 is taken \citep[see e.g.,][]{wade85}. The ratio
of the amplitude of the horizontal to the vertical motion, denoted by
$K$, is given by $K=GM/\omega^2 R^3$, where $M$ is the mass, $R$ the
radius and $\omega$ the angular pulsation frequency. With $M = 13.5
M_\odot$ and $R = 7.8 R_\odot$ \citep{decat02,heynderickx94} we obtain
\mbox{$K_1 = 0.133$}, \mbox{$K_2 = 0.132$} and \mbox{$K_3 = 0.121$}. We
varied the free parameters in the following way: the projected rotation
velocity, $v \sin i$, from 1 to 35~\kmps\  with a step 1~\kmps,  the
inclination angle of the star, $i$, from 1$^{\circ}$ to 90$^{\circ}$
with a step 1$^{\circ}$, and the line-profile width due to thermal
broadening, $\sigma$, from 1 to 20~\kmps\ with a step 1~\kmps.

The mode identification by means of the moment method gives a preference
to $m_1 = 2$ but we cannot rule firmly out $m_1 = 1$. For the mode with
frequency $f_3$ we cannot conclude anything, as is the case for the
photometric data \citep{shobbrook06}.

\subsubsection{The amplitude and phase variations across the line profile}
\label{sec:spec_modeid_ampphase}

Because the discriminant values of the best moment solutions are very
similar we need an additional check in order to safely conclude that
$(\ell_1,m_1) = (2,2)$. Our method is to visualise the behaviour of
observed amplitude and phase variations across the line profile compared
to theoretically computed ones for the best parameter sets given by the
moment method.

The observed amplitude and phase variations across the Si\,III 4553~\AA\
line are shown in Fig.~\ref{fig:amp_phase} for $f_1$, $f_2$ and $f_3$.
The theoretical distributions were computed from line profile time
series generated by means of Townsend's codes \citep{townsend97}, called
BRUCE and KYLIE. The line-profile variations as well as the amplitude
and phase variations were computed by considering the three modes
together. 

The amplitude distributions for $f_1$ computed for the best parameter
combinations for $(\ell_1,m_1) = (2,2)$ and $(\ell_1,m_1) = (2,1)$ given
by the moment method are shown in Fig.~\ref{fig:amp_phase_th}. By
comparing the observed (see Fig.~\ref{fig:amp_phase}) to the best
theoretical amplitude distributions (see Fig.~\ref{fig:amp_phase_th})
for the mode with frequency $f_1$, it becomes clear that $(\ell_1,m_1) =
(2,2)$. Indeed the behaviour of the theoretical amplitude distribution
differs from the observed one for all best solutions with $(\ell_1,m_1)
= (2,1)$ since the ``triple-humped'' character of those solutions are
absent in the observed amplitude distribution (see top left panel of
Fig.~\ref{fig:amp_phase}). All best parameter combinations with
$(\ell_1,m_1) = (2,2)$ mimic very well the observations. For the third
mode we unfortunately could not discriminate between the different
solutions. All we could derive from its phase behaviour in
Fig.~\ref{fig:amp_phase} is that $m_3 > 0$.

\subsection{Derivation of the stellar equatorial rotational velocity}
\label{sec:spec_vel}

\begin{figure}
\begin{center}
\includegraphics[width=85mm]{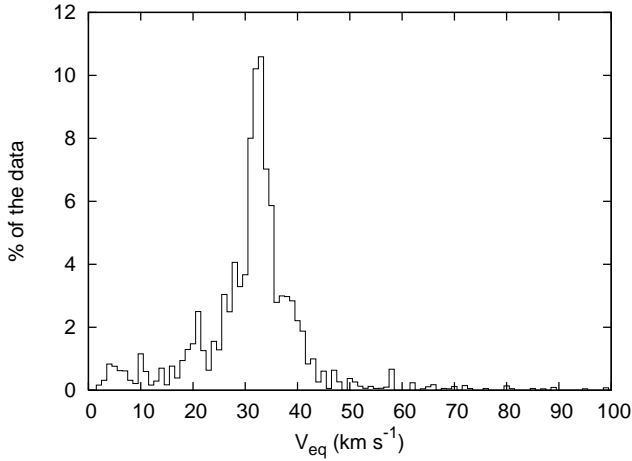}
\caption{
Histogram for the equatorial rotational velocity of the star.  
\label{fig:histo}
}
\end{center}
\end{figure}

Each solution given by the moment method indirectly gives a value for
the equatorial rotational velocity, since the inclination $i$ and the
projected equatorial velocity $v_{\Omega}=v_\mathrm{eq}\sin i$ are
estimated.  We made a histogram (see Fig.~\ref{fig:histo}) for
$v_\mathrm{eq}$ by considering only solutions with $(\ell_1, m_1)=(2,2)$
and by giving each equatorial rotational velocity $v_{\mathrm{eq},k}$
its appropriate weight $w_k=\Sigma_0/\Sigma_k$, where $\Sigma_0$ is the
discriminant value for the best solution. By calculating a weighted mean
and standard deviation of the data we got $v_\mathrm{eq} = 31 \pm
5~\kmps$ for $\beta$~Canis Majoris.  We also constructed a histogram for
the inclination angle in a similar way. We found a flat distribution in
the range $[0^\circ,90^\circ]$ so that we could not restrict the value
for $i$.  The moment method could consequently allow us to limit the
range for the couple $(v\sin i,i)$ but not $v\sin i$ or $i$ separately.

\section{Seismic interpretation}
\label{sec:model}

We have performed a thorough seismic analysis of the observed
frequencies by comparing them with frequencies obtained from theoretical
stellar models. Apart from the two identified frequencies, $f_1$ and
$f_2$, we have also taken into account the position of \bcma\ on the
\hrd\ in our modelling. The third unidentified frequency, $f_3$, is not
used in the modelling process.

\subsection{Global parameters}
\label{sec:model_global}

The effective temperature of \bcma\ has been quoted in the literature in
a range of values around $24000$~K. The extreme values are $\lteff =
4.37$ \citep{heynderickx94} and $\lteff = 4.45$ \citep{tian03}.
Recently \citet{morel06} made a detailed NLTE analysis of \bcma\ to find
$\lteff = 4.38 \pm 0.02$. We shall adopt a conservative range of $\lteff
= 4.40 \pm 0.04$ for the present work.

The luminosity can be determined from the Hipparcos parallax: $\pi =
6.53 \pm 0.66$~mas \citep{perryman97}. Assuming a bolometric correction
of $-2.29$~mag, this translates to a luminosity range of $\log L/\lsun =
4.41\pm 0.16$. However, the value of the bolometric correction might be
a major source of error in this calibration. We also find a range of
values in the literature, from $\log L/\lsun = 4.40$ \citep{stankov05}
to $\log L/\lsun = 4.79$ \citep{tian03}.  We adopt the range $\log
L/\lsun = 4.45\pm 0.20$, which covers most of these quoted values.

The metallicity of \bcma\ has been found to be $\mbyh=0.04\pm 0.10$ in
recent studies \citep{niemczura05}. We have adopted this range in our
models. Assuming a solar metallicity of $Z_\odot = 0.018$, this
translates into a range of metallicity for \bcma: $0.016 \leq Z \leq
0.023$. For a given metallicity, we have also varied the hydrogen
abundance slightly while exploring the parameter space for the models.

\subsection{Rotational splitting}
\label{sec:model_split}

Since the rotational velocity of \bcma\ has been found to be moderately
low, we do not expect any strong coupling between rotation and
oscillation frequencies. However, one cannot be certain if the interior
rotation velocity is higher or not. Indeed, evidence of differential
rotation of the core has been found in two other \bcep\ stars
\citep{dupret04,pamyatnykh04}. But in the absence of observations of a
rotational multiplet, or a definite estimate of the rotation velocity,
we do not have enough constraints to check if the internal rotation is
indeed uniform or not. As a first approximation, therefore, we have
assumed rigid rotation in the interior of the star. 

We constructed non-rotating stellar models and accounted for the effect
of rotation by adding the rotational splitting $\delta\nu_s = m\beta
v_\mathrm{eq}/R$ to the theoretical frequency, where $\beta$ represents
the Ledoux constant and depends on the rotational kernels.  The value of
$\beta$ is calculated for each mode from the eigenfunction.  The
rotational splitting is found to lie between $0.9~\muhz$ and
$1.3~\muhz$, depending on the model, the uncertainty being primarily due
to the rotational velocity, $v_\mathrm{eq}$. Since $\delta\nu_s$ depends
on $\beta$, instead of using a uniform value for $\delta\nu_s$, we have
calculated it for each particular mode of every model that we want to
compare with the observations. Thus, we calculate the theoretical value
of the rotationally split component ($\ell=2,m=2$) of the model
frequency: $\nu_\mathrm{model}(n,\ell=2,m=2) =
\nu_\mathrm{model}(n,\ell=2,m=0) + \delta\nu_s$ and match it with the
observed frequency $f_1$.

\subsection{Stellar models}
\label{sec:model_models}

We constructed a grid of stellar models between masses of $12\msun$ and
$16\msun$ using the CESAM evolutionary code \citep{morel97}. These
models used the OPAL equation of state \citep{rogers02} and OPAL opacity
tables \citep{iglesias96}, complemented by the low-temperature opacity
tables of \citet{alexander94}. Convection was described by the standard
mixing length theory \citep{henyey65} and nuclear reaction rates were
obtained from the NACRE compilation \citep{angulo99}. The frequencies of
oscillation were computed under the adiabatic approximation, using the
Aarhus pulsation package, ADIPLS \citep{jcd91}.

Our models adequately span the range of stellar parameters such as mass
($12 \leq M/\msun \leq 16$), chemical composition ($-0.06 \leq \mbyh
\leq 0.14$) and core overshoot ($0 \leq \dov \leq 0.3$, where \dov\
represents the extent of overshoot in terms of the local pressure scale
height). For each combination of these parameters we constructed models
spanning the main sequence phase of evolution. The frequencies of models
lying within or close to the error box on the \hrd\ were then compared
to the observed frequencies of \bcma.  The mixing length of convection
was not varied since it does not play a significant role in the models,
the outer convective envelope being extremely thin in such stars.
Diffusion and radiative levitation of elements were not incorporated in
the models.

\subsection{Matching the frequencies}
\label{sec:model_match}

Although the ($\ell,m$) values of two of the modes were identified from
their photometric and spectroscopic characteristics, their radial orders
cannot be determined from observations. Therefore, we have to account
for various possibilities for the radial order in the seismic modelling.
First we deal with the radial mode, $f_2$, by comparing it with the
radial frequencies of the stellar models. It becomes evident that the
radial mode must be either the fundamental mode ($\ell=0,n=1$) or the
first overtone ($\ell=0,n=2$) for the star to be at its prescribed
position on the \hrd. If $f_2$ were to be any higher order radial
frequency, then the radius of the star (hence the luminosity and \teff)
would become completely inconsistent with the estimates of \lteff\ and
$\log L/\lsun$ for \bcma\ obtained independently. Indeed, either of the
fundamental mode or the radial overtone has been observed in other
\bcep\ stars as well \citep[e.g.,][]{dupret04,ausseloos04,aerts06}. We
explored both possibilities by comparing the observed frequency, $f_2$
with the theoretical radial mode frequencies of our models.

\subsection{Radial mode -- fundamental or overtone?}
\label{sec:model_avcross}

As a first step, we matched $f_2$ to the radial fundamental mode of our
models. The frequency $f_1$ was then compared to different orders of
$\ell=2$ $g$- and $p$-modes (corrected for rotational splitting, as
explained in Sect.~\ref{sec:model_split}).  It turns out that for the
frequency range of interest, this mode being the first $g$-mode ($g_1$:
$\ell=2,n=-1$) is the only possible solution. The so-called $f$-mode
($\ell=2,n=0$) is too high and the second $g$-mode ($g_2$:
$\ell=2,n=-2$) is too low in value compared to $f_1 - \delta\nu_s$.

However, although it is easy to find a number of models with different
stellar parameters whose radial fundamental mode matches with $f_2$,
none of these models match the frequency $f_1$. The central problem lies
in the fact that the $g_1$ mode frequency of any given model is too
close to the radial fundamental mode, compared to the distance between
them as indicated from the observations of $f_1$ and $f_2$. The only
other nearby $\ell=2$ mode, the $g_2$ mode, is far too distant from the
radial fundamental.  In other words, the ``separation'' between the
radial fundamental mode and the $\ell=2$ modes in the models is either
much smaller (for the $g_1$ mode) or much larger (for the $g_2$ mode)
than the observed value, $f_2-(f_1 - \delta\nu_s)$. What we require to
match both frequencies is an intermediate value ($\sim 1.2~\muhz$) of
the separation between the radial mode and the $\ell=2$ mode.

\begin{figure}
\begin{center}
\includegraphics[width=85mm]{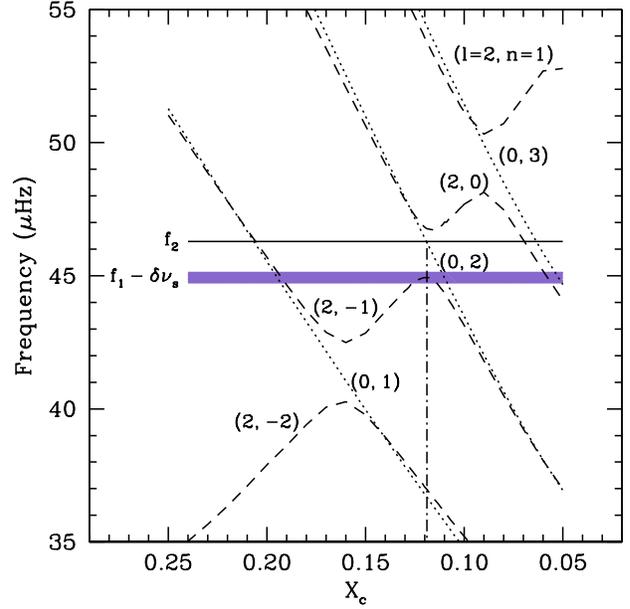}
\caption{
The variation of the frequency of $\ell=0$ and $\ell=2$ modes of a
$13\msun$ stellar model with evolution (characterised by the central
hydrogen abundance, \xc) is shown. The dotted lines indicate the
$\ell=0$ modes, while the dashed lines show the $\ell=2$ modes. Each
mode is labelled by its degree and radial order ($\ell, n$). The
horizontal solid line and the shaded strip show the observed radial mode
($f_2$) and the splitting-corrected $\ell=2$ mode ($f_1 - \delta\nu_s$)
respectively. The vertical dot-dashed line shows a possible solution
where the theoretical frequencies match the respective observed
frequencies.
\label{fig:avcross}
}
\end{center}
\end{figure}

The clue to our problem lies in recognising the fact that such a small
(but not too small) difference between the $\ell=0$ and $\ell=2$ modes
can only occur in case of an avoided crossing. As long as the modes
follow the regular smooth separation patterns, we will never find a
solution where the $\ell=0$ and $\ell=2$ modes have the appropriate
separation as required by the observations. This is illustrated in
Fig.~\ref{fig:avcross}. In this figure, we show how the $\ell=0$ and
$\ell=2$ modes vary with evolution of a given star. For the purpose of
illustration, we have assumed an average value of the rotational
splitting (with an error margin) and corrected the observed frequency,
$f_1$, to compare it with the theoretical frequency. As the star
evolves, the density gradient at the edge of the shrinking convective
core increases. This has the effect of rapidly increasing the
frequencies of the $g$-modes, which causes successive ``bumping'' of the
modes. For example, as shown in Fig.~\ref{fig:avcross}, the $g_2$ mode
of the $\ell=2$ degree first increases (for $\xc \geq 0.16$), until it
bumps into the $g_1$ mode (around $\xc \simeq 0.16$), creating an
avoided crossing between these two modes at that age. With further
evolution, the $g_1$ mode increases in value until it creates another
avoided crossing with the $f$-mode (around $\xc \simeq 0.12$).
Progressively, higher order modes are bumped as the star evolves. An
excellent discussion of such mode bumping is provided by
\citet{aizenman77}.

We find that as long as we are constrained by the position of \bcma\ on
the \hrd, the required separation between the $\ell=0$ and $\ell=2$
modes ($\sim 1.2~\muhz$ around a frequency value of $45~\muhz$) cannot
be obtained if the radial mode is the fundamental mode. Even for an
avoided crossing, the appropriate separation between the radial
fundamental and the $\ell=2,n=-2$ bumped up mode occurs at a much lower
frequency than observed.  Fig.~\ref{fig:avcross}, which represents a
typical model at the appropriate position on the \hrd, confirms this.
However, a solution is indeed plausible if we consider the observed
radial mode to be the first overtone instead. In that case, the avoided
crossing between the $\ell=2$ $g_1$ and $f$-modes bumps the $g_1$ close
to the $\ell=0,n=2$ mode. We now, therefore, explore the possibility of
the radial mode being the first overtone mode.

By considering the radial mode as the first overtone ($\ell=0,n=2$), we
do indeed find a number of models with different stellar parameters that
match the observed frequencies. Specifically, $f_2$ is matched with the
$\ell=0,n=2$ mode, and $f_1$ with the $\ell=2,n=-1$ mode, after
correcting for the rotational splitting. We have also checked the
excitation rates of these two frequencies by the nonadiabatic
oscillation code, MAD \citep{dupret01} and found both of them to be
excited for all the models which fit the data.

\begin{figure*}
\begin{center}
\resizebox{56mm}{!}{\includegraphics{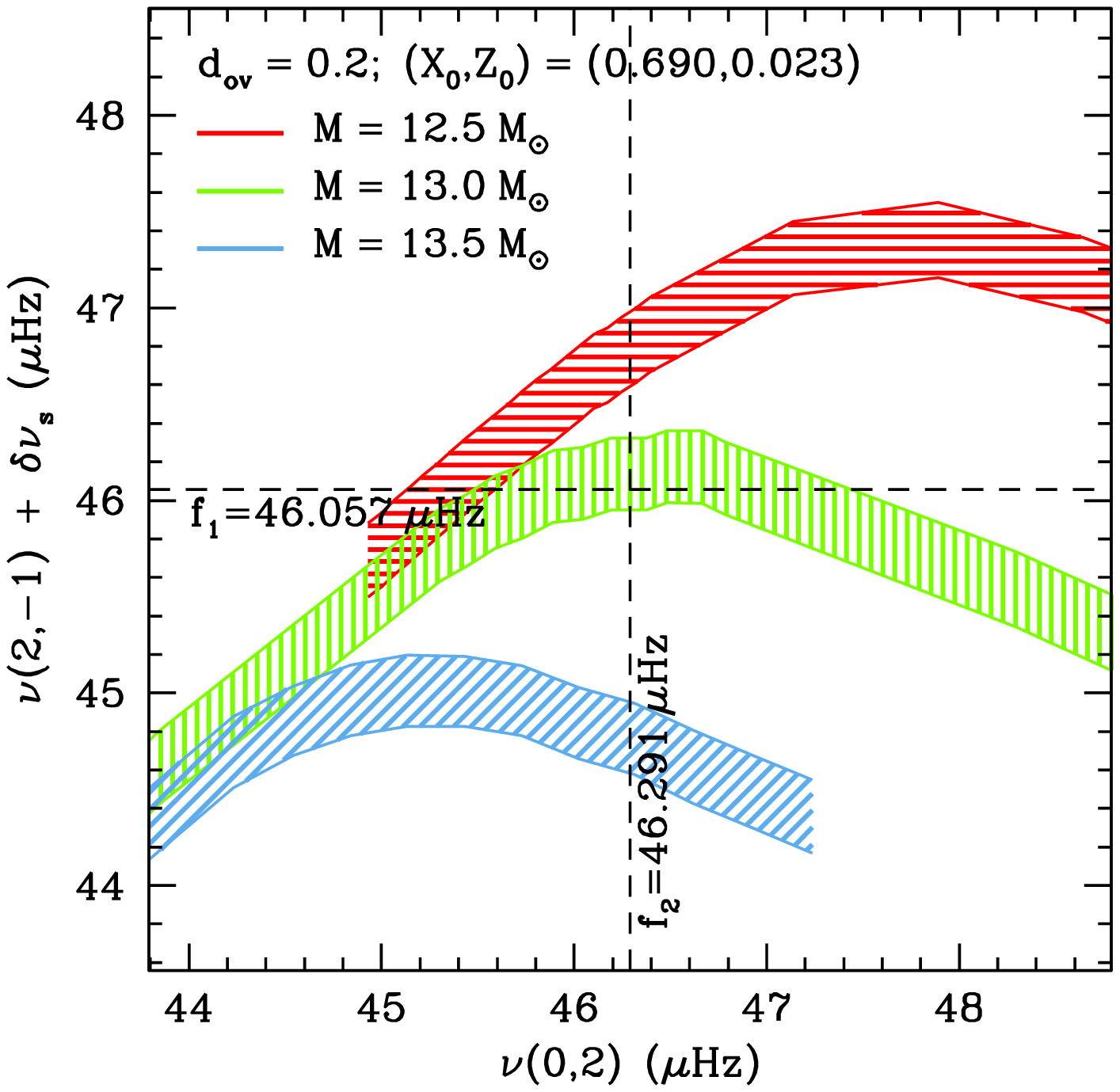}}\hfil
\resizebox{56mm}{!}{\includegraphics{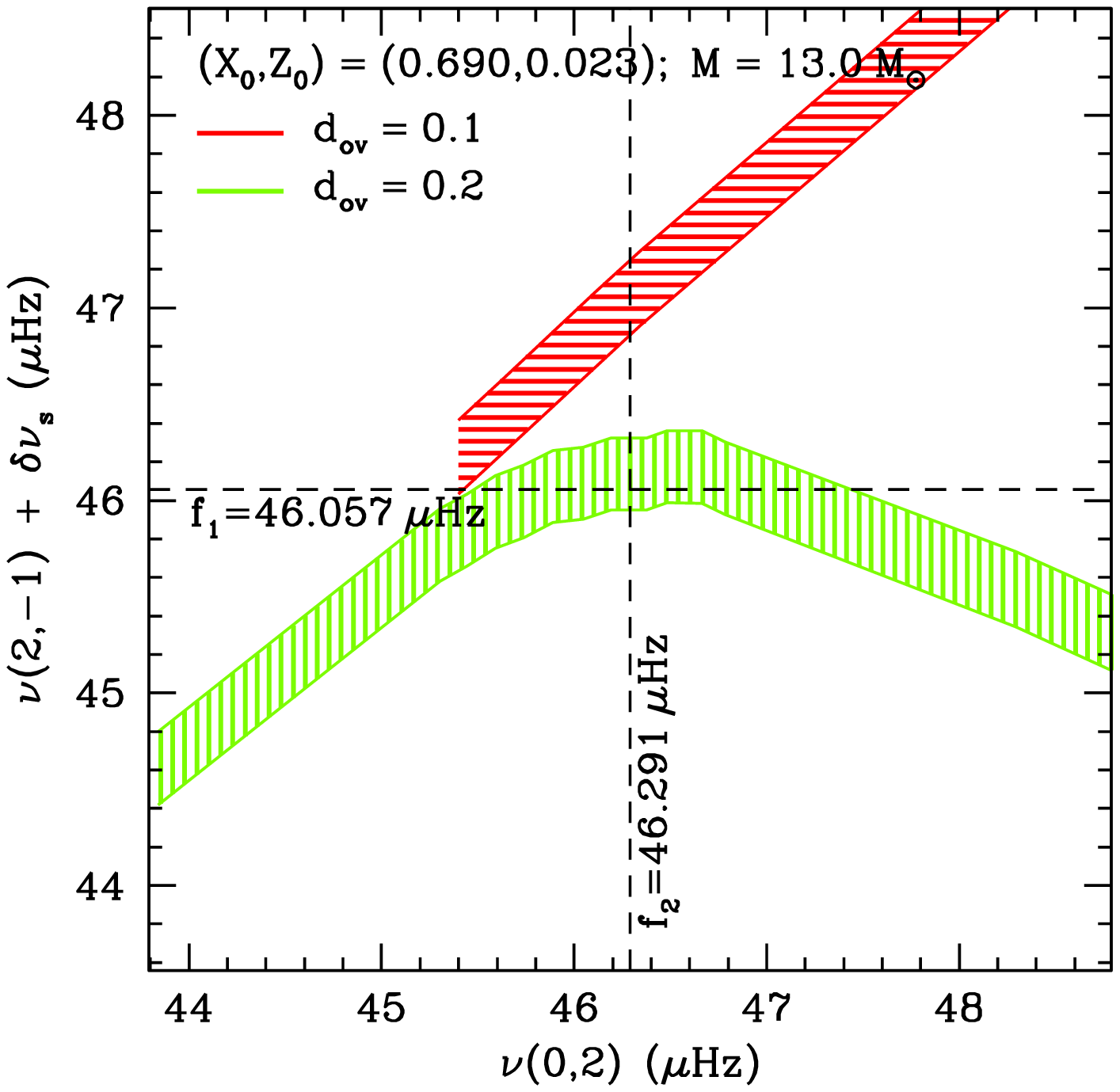}}\hfil
\resizebox{56mm}{!}{\includegraphics{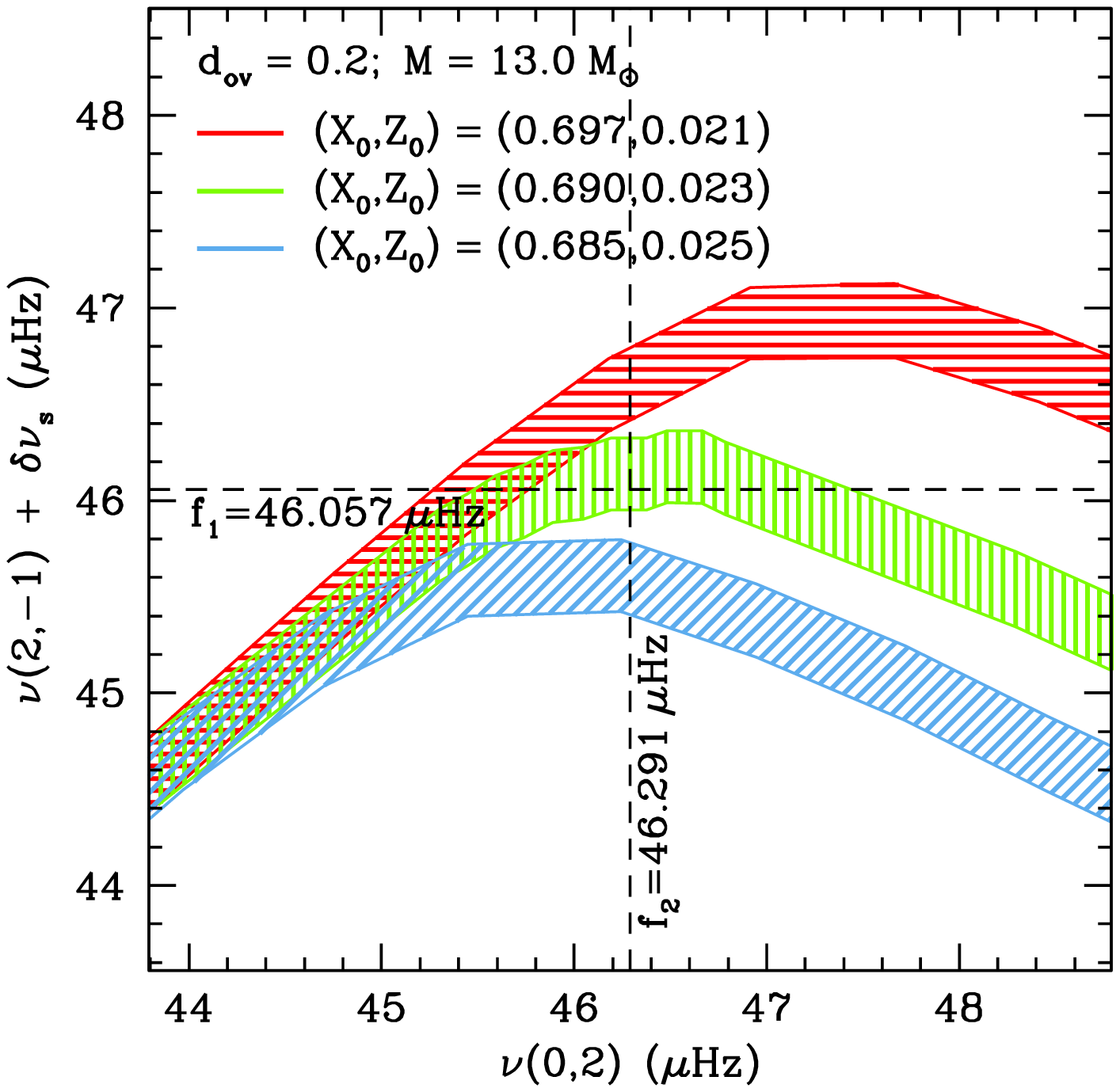}}
\end{center}
\caption{
The variation of the model frequencies with three stellar parameters --
$M$, \dov\ and $Z_0$ -- are shown. In each panel, two of the parameters
are fixed to clearly bring out the effect of the remaining parameter on
the frequencies. The $\ell=0,n=2$ frequencies are plotted as abcissae,
while the $\ell=2,n=-1$ frequency, corrected for rotational splitting,
are plotted as ordinates. The dashed lines indicate the two observed
frequencies. The point where the two dashed lines intersect indicate the
perfect match of the desired model. Each shaded band represents an
evolutionary track. The width of the band along the $y$-axis indicates
the uncertainty due to the rotational splitting. A full-colour version
of this figure is available in the online edition.
\label{fig:param}
}
\end{figure*}

A similar solution with the radial mode being the second overtone
($\ell=0,n=3$) and the $\ell=2$ mode being the bumped-up $f$-mode is
again ruled out because the relevant avoided crossing occurs at a higher
frequency than observed.

The common feature of all the possible solutions is the fact that the
$\ell=2,n=-1$ mode is an avoided crossing with the $f$-mode.  This helps
to constrain the stellar parameters a great deal, especially the age (or
\xc) of the model. The requirement that the star must be at the precise
age for a specific avoided  crossing to occur, is indeed a very strict
one.  Nevertheless, since we have only two frequencies to constrain our
model with, we do find multiple combinations of the stellar parameters
where such a solution exists. We have investigated the limits of the
stellar parameters that can be constrained with the observed frequencies
(see next section). Although the limits on effective temperature and
luminosity were not explicitly used in determining the best models, all
of them do lie well within the errorbox on the \hrd\ adopted in
Sect.~\ref{sec:model_global}.

\subsection{Limits on stellar parameters}
\label{sec:model_param}

We have, in principle, four major stellar parameters to tune our models
with -- $M$, \dov, $X_0$ and $Z_0$. For each appropriate combination of
these parameters, the age (in terms of \xc) is automatically chosen by
the closest match of the frequencies. The initial hydrogen abundance,
$X_0$, must, however, be linked somewhat to the metallicity, $Z_0$; we
have varied $X_0$ only within permissible limits for a particular choice
of $Z_0$ so that \mbyh\ remains within the bounds quoted in
Sect.~\ref{sec:model_global}. In Fig.~\ref{fig:param}, we have shown how
the relevant frequencies of the models, $\nu(\ell=0,n=2)$ and
$\nu(\ell=2,n=-1)$, vary as each of the stellar parameters are changed.
In each panel, only one parameter is changed at a time, keeping the
others constant. Each band is, in fact, an evolutionary track -- the
sense of evolution being from the right to the left (decreasing radial
frequency). The width of the shaded bands along the $y$-axis reflects
the uncertainty due to the rotational splitting correction. The tracks
are truncated beyond the ages where the models lie outside the error box
on the \hrd.  Our best models lie near the centre of each plot, where
both the model frequencies match the observed ones. 

\begin{figure}
\begin{center}
\resizebox{0.9\hsize}{!}{\includegraphics{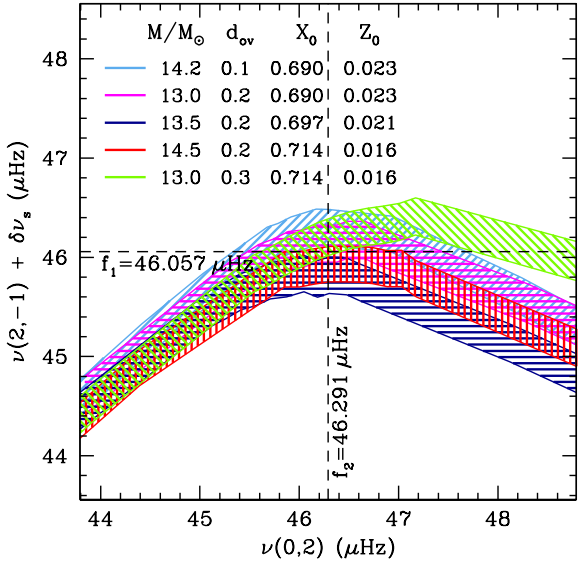}}
\caption{
Similar to Fig.~\ref{fig:param}, except that multiple parameters are
varied at a time to obtain a good match between the theoretical
frequencies and the observed ones. Selected models with limiting values
of overshoot and metallicity are shown only. A full-colour version of
this figure is available in the online edition.
\label{fig:bestmodels}
}
\end{center}
\end{figure}

In Fig.~\ref{fig:bestmodels} we vary multiple parameters simultaneously
to obtain a good match of the model frequencies with the observed ones.
This illustrates that several solutions are possible in the
multi-dimensional parameter space.  Actually, we have shown only the
models with the extreme limits of the stellar parameters for which a
solution can still be found. These are only indicative of the trend of
the frequencies of the models, and several other solutions are possible
when the parameters (especially overshoot and metallicity) are varied
within their bounds. Table~\ref{tab:bestmodels} lists the physical
parameters for these selected models.

We find that no solutions are possible in the absence of core overshoot.
Even with a small amount of overshoot ($\dov =0.1$), we need to have a
higher mass to obtain a solution. The higher stellar mass helps to
increase the mass of the convective core to offset the low overshoot.
Most of our best fit models use $\dov=0.2$. Higher overshoot ($\dov =
0.3$) models can reproduce the frequencies only for low metallicity and
proportionally lower helium content, again suggesting a trade-off
between overshoot and helium content to maintain a balance of the core
size. This indicates that the size of the convective core plays a
crucial role in determining the frequencies. Indeed, all our best fit
models, including those with low and high overshoot have fractional
convective core mass of $24$--$28\%$.

\begin{table*}
\begin{center}
\caption{
Physical parameters of representative stellar models that match the 
observed frequencies. 
\label{tab:bestmodels}
}
\begin{tabular}{ccccccccc}
\hline
\hline
$M/\msun$ & \xc & $X_0$ & $Z_0$ & \dov & \lteff & $\log L/\lsun$ &
$R/\rsun$ & Age(Myr) \\
\hline
14.2 & 0.105 & 0.690 & 0.023 & 0.1 & 4.378 & 4.530 & 10.67 & 10.95 \\
13.0 & 0.127 & 0.690 & 0.023 & 0.2 & 4.364 & 4.440 & 10.35 & 13.04 \\
13.5 & 0.128 & 0.697 & 0.021 & 0.2 & 4.373 & 4.488 & 10.50 & 12.48 \\
14.5 & 0.123 & 0.714 & 0.016 & 0.2 & 4.391 & 4.582 & 10.75 & 11.75 \\
13.0 & 0.134 & 0.714 & 0.016 & 0.3 & 4.370 & 4.471 & 10.40 & 14.49 \\
\hline
\end{tabular}
\end{center}
\end{table*}

The mass of our best fit models are mostly limited to $13.5 \pm 0.5
\msun$, for $\dov = 0.2$. The mass could be higher (up to $14.5\msun$)
if either the metallicity is low ($\mbyh=-0.06$) or the overshoot is low
($\dov=0.1$). The central hydrogen abundance of the best models are
limited to the range $0.128 \geq \xc \geq 0.123$, for overshoot values
of $\dov=0.2$. As expected, the low overshoot models have younger ages
and lower \xc. The situation is opposite for high overshoot models.

\section{Conclusions}
\label{sec:concl}

$\beta$~Canis Majoris is one of the \bcep\ stars whose variability has
been observed and analysed for one century. It was discovered that this
star pulsates with three frequencies rather low in comparison with other
known stars of its type. For this reason \bcma\ is an important target
for asteroseismology purposes.  However, so far, no definite mode
identification had been achieved for this star so that no modelling
could be attempted. Our aim was to increase the number of known
pulsating frequencies and mostly to provide a unique identification of
the modes of \bcma.

Our study was based on 452 ground-based high-resolution high S/N
spectroscopic measurements spread over 4.5 years. We used the Si\,III
4553~\AA\ line to derive the pulsation characteristics of \bcma.  Our
dataset unfortunately suffers from strong aliasing but the three
established frequencies of the star were confirmed in the first three
velocity moments of the line and in the spectra themselves. They are
$f_1 = 3.9793~\cpd$ ($f_1 = 46.057~\muhz$), $f_2 = 3.9995~\cpd$ ($f_2 =
46.291~\muhz$) and $f_3 = 4.1832~\cpd$ ($f_3 = 48.417~\muhz$).
Unfortunately no new frequencies were discovered neither in our
spectroscopic observations nor in the recent multisite photometric
measurements led by \citet{shobbrook06}. 

The important result of the combination of both intensive campaigns is
an identification of the two main modes of \bcma, which is a strong
constraint for further asteroseismic modelling of the star. The
photometric identification by \citet{shobbrook06} yielded $\ell_1 = 2$
and $\ell_2 = 0$. Our spectroscopic data could corroborate that the mode
with $f_2$ is radial. We adopted the photometric identification of
$\ell_1$ and spectroscopic techniques allowed us to derive the $m$-value
of the main mode. The application of the moment method gave a preference
to $m_1 = 2$. Because moment solutions could not definitely exclude $m_1
= 1$ we made use of the behaviour of the amplitude distributions across
the line profile for the best parameter sets given by the moment method.
In this way we could conclude without any doubt that $(\ell_1,m_1) =
(2,2)$. For the third mode nothing could be concluded. In addition we
derived a stellar equatorial rotational velocity of $31 \pm 5 \kmps$.

The definite identification of two of the observed modes and a much
improved estimate of the rotation velocity of \bcma\ allowed us to
attempt the first seismic modelling of this star. Although it is not
realistic to hope for a unique model to fit just two frequencies, we
have thoroughly explored the stellar parameter space to derive
reasonable constraints for the mass, age, and core overshoot.  The most
significant aspect of the seismic analysis is the fact that we could
assert that the non-radial mode, $f_1$, is close to an avoided crossing.
This implies a very strong constraint on the stellar parameters,
especially the age of the star. At the same time, it rules out the
possibility of the radial mode, $f_2$, being the fundamental mode. This
makes \bcma\ one more \bcep\ star known to have a dominant radial
overtone mode of pulsation \citep[cf.][]{aerts06}. 

Our best fit models indicate that \bcma\ has a mass of $13.5\pm 0.5
\msun$, an age of $12.4\pm 0.7$~Myr ($\xc = 0.126\pm 0.003$) and core
overshoot of $\dov = 0.20\pm 0.05$. No satisfactory model can be found
if core overshoot is absent.  A small overshoot parameter is possible
only for a higher mass along with high metallicity (and proportionally
higher helium content). On the other hand, higher core overshoot is
required if the star is metal-poor ($\mbyh < -0.05$).  However,
\citet{morel06} found the composition of \bcma\ not much different from
that of the Sun, making such possibilities unlikely.  Therefore, it is
safe to conclude that the models with $\dov=0.20\pm 0.05$ are the most
likely ones.  If the chemical composition of \bcma\ could be known to
higher accuracy independently, one would be able to constrain the other
parameters even more.  All the solutions turn out to have effective
temperatures close to the cooler edge of the adopted  errorbox on the
\hrd. This is consistent with the recent estimates of \teff\
\citep[e.g.,][]{morel06}. We note that we had deliberately chosen a very
conservative errorbox for effective temperature and luminosity; a
stricter limit on these parameters would rule out some of our possible
models. 

In retrospect, one can also try to identify the mode of oscillation for
the third frequency, $f_3$, by comparing it with the theoretical model
frequencies. In this comparison, we allowed for different rotational
splitting values, $m_3$, for each non-radial mode with the restriction
$m_3>0$, as we found in Sect.~\ref{sec:spec}. This leads us to only one
possibility for $f_3$, for all the models which match $f_1$ and $f_2$:
$\ell_3 = 3, m_3 = 2, n_3=-1$. We hope that this identification of $f_3$
can be checked through future observations.  However, we cannot place
further constraints on the models at this stage using this frequency;
all the models with stellar parameters in the range restricted by the
first two frequencies also match the third frequency with the
identification given above within the uncertainty associated with the
rotational velocity. One would need a more precise estimate of the
rotational velocity to distinguish between these models.
  
The rotation period may be calculated from our estimate of the
equatorial velocity (Sect.~\ref{sec:spec_vel}) and the radius of our
best models which indeed lie in a narrow range
(Table~\ref{tab:bestmodels}). We estimate the rotation period to be
$18.6 \pm 3.3$~days, which indicates that \bcma\ is indeed a slow
rotator; therefore, our assumption in neglecting higher order terms of
the rotation velocity while calculating the frequency splitting stands
justified.  

Despite the knowledge of only two frequencies for this star, the
occurrence of the avoided crossing goes a long way towards constraining
most of the stellar parameters. While one cannot expect to be so lucky
for every star, we have shown that the identification of an avoided
crossing might help us to extract a lot more information about the star
than any normal mode.

\begin{acknowledgements}
We thank all the observers from the Institute of Astrophysics of the
University of Leuven who gathered the spectroscopic data used in the
current paper. The authors are supported by the Research Council of
Leuven University under grant GOA/2003/04.  
\end{acknowledgements}


\end{document}